# Shape sensitive Raman scattering from Nano-particles

S.P. Apell[1,2], G. Mukhopadhyay[1,3], Tomasz Antosiewicz[1,4] and J Aizpurua[2]

1 Department of Physics and Gothenburg Physics Centre, Chalmers University of Technology, SE-412 96 Göteborg, Sweden.

2 Materials Physics Center CSIC-UPV/ EHU and Donostia International Physics Center DIPC, Paseo Manuel de Lardizabal 5, Donostia-San Sebastian 20018, Spain

3 Physics Department, Indian Institute of Technology, Powai, Mumbay 400076, India

4 Centre of New Technologies, University of Warsaw, Żwirki i Wigury 93, 02- 089 Warsaw, Poland.

## Abstract

We investigate the interplay of shape changes and localized surface plasmons in small metal particles with the potential of a large enhancement of the Raman signal from the particles own vibrations. The framework is a geometrical one where we study the change in geometric factors during the vibrational movement. The resulting cross-section is found to be of a detectable order of magnitude however much smaller than the elastic cross-section.

# Raman scattering from single metal particles

Raman scattering is a powerful tool to understand the interplay between vibrational and electronic degrees of freedom in a system. For many years this has been a very useful tool in molecular physics now bordering the single molecule detection limit [1]. From the scattered light, which is vibrationally shifted with respect to the incoming light, we can infer important properties of the system at hand. One of the major goals in this area has for many years been to be able to monitor the Raman scattering using individual molecules to probe a particular local environment. A ubiquitous problem in this case has been to get enough field strength to pick up an informative signal; often handicapped by signals averaged over many molecules and hence many local environments. Considerable progress has therefore been made in this area using the excitation of local electromagnetic resonances in order to enhance both the incoming and the scattered radiation fields. In particular small metal particles are very interesting from this point of view with their characteristic surface plasmon modes which are tunable to a particular energy by simply changing their shape, size or local dielectric environment. With the advent of new fabrication and detection techniques, allowing in principle for an ability to pick up the single particle Mie spectrum, one should also be able to pick up the inelastic Mie scattering part; especially if it can be enhanced by local surface plasmon/polariton resonances.

Raman scattering from surfaces with our without molecules has a long history [2-8]. It was followed up focussing on details in single particle or molecular vibrations in [9-15] and was preceeded by collodial studies in [16-18]. A special direction of research has been the vibrations of elongated particles [19-24]. In recent years [25-34] one has focussed on more and more elaborated systems such as arrays of particles, higher order multipoles and the influence of the surrounding medium. In this context we should remind the reader about a development in cluster physics [35-37] approaching from below in size the development indicated above.

Typical vibrational energies are in the range of a few meV for nanosized particles. Classic elasticity theory is often used where the frequency of vibration is related to the corresponding sound velocity and in the simplest models inversely proportional to size of the of the particle. [38]. For a spherical particle one distinguishes between spheroidal modes (purely radial displacements) and torsional modes (purely tangential displacements). Duval early on pointed out [39,40] that L=0,2 spheriodal modes are Raman active but all torsional ones are Raman inactive. G. Bachelier and A. Mlayah [41] investigated in what way a vibration of a particle could couple to surface plasmon polaritons and found two contributions. One is a modulation of the interband susceptibility via deformation potential changes due to purely radial vibrations. The other is a more surface localized effect where quadrupolar vibrations modulate the surface polarization. A change of particle shape gives a change in the surface polariton polarization vectors. Pure transverse and pure radial vibrations are however not active since such modes do not change the shape of the particle and hence to lowest order only quadrupolar spheroidal vibrations contribute.

Several studies have been performed on rods where a breathing mode dominates; a pure radial mechanical deformation of the rod with no change in length. Also extensional modes induced by laser heating has been demonstrated where the rod waist oscillates out and in changing both length and width. If the rod has indentations they can act as acoustic hotspots with both surface plasmons and strong shape deformations in the same place. [42]. For both rods and particles a thermal expansion can trigger anisotropic shape oscillations implying the ionic background is moving, hence interacting with the electronic system and probing its responsiveness of setting up local plasma oscillations.

In what follows we investigate the Raman scattering properties of individual metal particles with an emphasis on shape effects. We make realistic estimates of anticipated cross-sections to validate the feasibility of experimentally detecting them. It turns out that a type of quadrupolar geometric factors together with the more well-known dipolar geometric factors completely characterize the shape related part of the vibrational response. This in turn can get greatly enhanced provided one can meet the condition to set up a local surface plasmon excitation in the particle. We use a simple classical electromagnetic field description throughout the paper.

## Raman cross-section

For a polarizable object irradiated with light of frequency $\omega_{inc}$ there will be an induced dipole moment $\vec{\mu}$ proportional to the amplitude $\vec{E}$ of the total electric field acting:

$$\mu_i(t) = \sum_j \alpha_{ij} \cdot E_j \cos \omega_{inc} t ,  \quad (1)$$

and *i,j* denote appropriate coordinates. The "constant" of proportionality, $\alpha_{ij}$, is the polarizability tensor of the particle and contains all its symmetry properties. If the particle performs an oscillatory motion, in our case an acoustic vibration with normal mode frequency $\omega_{nom}$, we may write to lowest order

$$\alpha_{ij}(t) = \alpha_{ij}^0 + \sum_k \left.\frac{\partial \alpha_{ij}}{\partial q_k}\right|_o q_k \cos \omega_{nom} t + ... . \quad (2)$$

$\alpha_{ij}^o$ is the polarizability tensor for the equilibrium situation. $q_k$ is a normal mode coordinate introducing an index *k* pertaining to the polarization of the normal mode. There is of course a whole set of selection rules depending on the symmetry of the mode *k*, the field acting as well as different types of normal modes. However we address these issues when pertinent further down. Combining Eqs. (1) and (2), and for illustrational purposes restricting to a situation where the polarizability tensor is diagonal (but not necessarily isotropic), we have:

$$\mu_i(t) = \alpha_{ii}^0 E_i \cos \omega_{inc} t + \frac{1}{2} \sum_k E_i \left. \frac{\partial \alpha_{ii}}{\partial q_k} \right|_o q_k [\cos(\omega_{inc} - \omega_{nom})t + \cos(\omega_{inc} + \omega_{nom})].... \quad (3)$$

The first term corresponds to the standard elastic Rayleigh scattering from an oscillating dipole (same frequency as $\omega_{inc}$) and the second term gives rise to Raman scattering at the so called Stokes and anti-Stokes frequencies with a cross-section $\sigma_R$ being proportional to the square of the change in the polarizability as the particle is vibrating. For our purposes we use the simplest possible form for the cross-section according to [43] (c being the velocity of light):

$$\sigma_R \cong \left(\frac{\omega_{inc}}{c}\right)^4 \left| \sum_k q_k \left. \frac{\partial \alpha_{ii}}{\partial q_k} \right|_o \right|^2, \quad (4)$$

neglecting any shift of the incoming and outgoing frequencies because of the vibrational mode.

### Particle polarizability

In order to proceed we now introduce a generic polarizability for our particle which illustrates in a realistic way how shape and also the size of the particle influences the Raman cross-section. To this order we use the dipolar polarizability for an ellipsoidal particle (p) of volume V in a surrounding medium (d) [44]:

$$\alpha_{ii}(\omega, L) = V \varepsilon_d \frac{\varepsilon_p(\omega) - \varepsilon_d(\omega)}{\varepsilon_d(\omega) + L_i [\varepsilon_p(\omega) - \varepsilon_d(\omega)]} \quad (5)$$

where $\varepsilon_p / \varepsilon_d$ is the dielectric permeability of the particle and surrounding medium respectively. For the simplest order of magnitude estimates we will use $\varepsilon_d = \varepsilon_0$, the free space value, and for $\varepsilon_p$ the simplest possible free-electron like Drude form for a metallic particle.

$L_i$ in the denominator of Eq.(5) is a geometric factor often mistakenly called depolarization factor (see page 147 of Bohren and Huffman [44]) and corresponding to the direction *i* of the applied field. It depends only on the shape of the particle [44,45]. In general $L_1 + L_2 + L_3 = 1$ and they can be evaluated analytically for the case of spheroids (prolate: $a_1 = a_2 < a_3$; oblate $a_1 = a_2 > a_3$) where $a_1 / a_2 / a_3$ are the lengths of the semi-axes of the ellipsoid along *x*, *y* and *z* respectively. For what comes below we only need $L_3$ ($L_1 = L_2 = (1 - L_3)/2$):

$$L_3 \equiv \frac{a_1 a_2 a_3}{2} \int_0^\infty ds \, \frac{1}{f(s)(a_3^2 + s))} = \frac{e^2 - 1}{e^2}[1 + \frac{1}{2e} \ln \frac{1-e}{1+e}] \text{ when } a_1 = a_2, \qquad 6)$$

defining $e^2 = 1 - a_1^2 / a_3^2$ and $f(s) \equiv \sqrt{(a_1^2 + s)(a_2^2 + s)(a_3^2 + s)}$. In the spherical limit (*e*=0) $L_3 = 1/3$ and it goes to zero when $e \to 1$ (needle). Oblate expressions can be obtained in a straightforward manner from the literature [44].

When the particle is executing a vibrational motion (whose basic mechanisms we will not consider in detail in this paper) it has the potential, depending on symmetry, to influence both the size and shape of the particle. This in turn will change the volume as well as the geometric factors involved in the expression for the polarizability in Eq. (5). Since we are mainly considering a resonant Raman scattering situation to have as large effect as possible shape effects (*L*) dominate size effects (*V*) as shown and discussed further in the Appendix. There we also address three effects which go beyond Eq.(5) above: radiation damping, dynamic depolarization and change in volume influencing the plasma frequency of the particle while we at this stage continue to analyze the simpler situation to see if the resulting cross-section is large enough to be detect in a reasonable experimental set-up.

We want to find how the polarizability of the particle changes with respect to a vibrationally induced deformation of the particle, in a particular direction. We therefore rewrite according to:

$$\left. \frac{\partial \alpha_{ii}}{\partial q_j} \right|_0 = \frac{\partial \alpha_{ii}}{\partial L_i} \frac{\partial L_i}{\partial a_j} \left. \frac{\partial a_j}{\partial q_j} \right|_o . \qquad (7)$$

It is straightforward to find $\partial \alpha_{ii} / \partial L_i = -(\alpha_{ii}^2)/V$. For the last derivative in Eq. (7) we use the following lowest order Taylor expansion for the semi axis *j*:

$$a_j \equiv a_j^o + q_j \frac{\partial a_j}{\partial q_j}\bigg|_o , \tag{8}$$

only allowing for a change in direction *i* from a normal coordinate change in direction *j*. Introducing a relative displacement $\delta_j \equiv (a_j - a_j^o)/a_j^o$ and a dimensionless shape change tensor:

$$\Gamma_{ij} \equiv a_j^o \frac{\partial L_i}{\partial a_j} , \tag{9}$$

we can write for our Raman cross-section:

$$\sigma_R \cong \left(\frac{\omega_{inc}}{c}\right)^4 \left|\sum_k q_k \frac{\partial \alpha_{ii}}{\partial q_k}\bigg|_o\right|^2 = \left(\frac{\omega_{inc}}{c}\right)^4 \left|\frac{1}{V} \alpha_{ii}^2 \sum_j \Gamma_{ij} \delta_j\right|^2 . \tag{10}$$

The implication of the findings above is that if we have incoming light close to a surface collective electronic resonance of the particle, we anticipate an enormous enhancement provided the resonance is reasonably sharp. A more detailed treatment gives that "one" polarizability should be evaluated at the incoming photon energy and the other "one" at the vibrationally shifted energy, potentially yielding a smaller overall enhancement in general. Now most surface plasmons are much broader than the few meV of the vibrational excitation for the situation at hand, so we do not see this as any major weakening of the possible enhancement.

### Shape factor for Raman scattering

We now focus on the shape-change tensor $\Gamma_{ij}$ as defined above in Eq.(9), i.e. by taking the relevant derivatives of the geometric factor $L_i$ pertaining to the direction *i* with respect to a change of particle dimension in the direction *j*. In order to facilitate this we first need to introduce a dimensionless factor $L_{ij}$ carrying the quadrupolar character of the distortion giving rise to the Raman scattering. Following the structure of $L_i$ we define $L_{ij}$ as:

$$L_{ij} = \frac{a_1 a_2 a_3 a_i^2}{2} \int_0^\infty \frac{ds}{f(s)(a_i^2+s)(a_j^2+s)} , \quad f(s) = \sqrt{(a_1^2+s)(a_2^2+s)(a_3^2+s)} \tag{12}$$

We will call $L_{ij}$ the geometric distorsion tensor. The symmetry of the problem ($a_1 = a_2$) makes it enough to know $L_{11}$ $L_{13}$ and $L_{33}$. Now we have that $L_{12} = L_{21} = L_{22} = L_{11}$ and $L_{31} = L_{32} = L_{13}/(1-e^2) = L_{23}/(1-e^2)$. $e$ was defined in connection with Eq.(6). However like the $L_i$'s, $L_{ij}$ fulfil sum-rules [46] so $L_{11} = (1 - L_{13})/4$ and $L_{33} = (1 - 2L_{13}/(1-e^2))/3$ which means it is enough to know $L_{13}$ to generate all the other $L_{ij}$'s. In fact using yet another sum-rule $4L_{11} + L_{13}/(1-e^2) = 3L_1 = 3(1-L_3)/2$ we can even generate $L_{13}$ from $L_3$ alone:

$$L_{13} = \frac{1-e^2}{2e^2}(1-3L_3) = \frac{1-e^2}{2e^4}[3 - 2e^2 + \frac{3}{2e}(1-e^2)\ln\frac{1-e}{1+e}] \tag{13}$$

$L_{13}$ starts out from *1/5* in the spherical limit (e=0) and goes to 0 in the needle limit (e=1). For most of the shape range it has very little variation. The same is true for the other $L_{ij}$ which are also of about the same order of magnitude.

From the definition of $\Gamma_{ij}$ in Eq.(9), combined with $L_1 + L_2 + L_3 = 1$, it follows immediately that $\sum_i \Gamma_{ij} = 0$. This together with the general symmetry properties at hand and the sum-rules implies that $\Gamma_{22} = \Gamma_{11}$, $\Gamma_{12} = \Gamma_{21}$, $\Gamma_{32} = \Gamma_{31} = -(\Gamma_{11} + \Gamma_{12})$, $-\tfrac{1}{2}\Gamma_{33} = \Gamma_{23} = \Gamma_{13}$. Hence it is enough to calculate $\Gamma_{13}$. Performing the derivative in Eq.(9) we get:

$$\Gamma_{13} = L_1 - L_{31} = \frac{3(e^2-1)}{2e^4}[1 + \frac{(1-e^2/3)}{2e}\ln\frac{1-e}{1+e}]. \tag{14}$$

It starts out from *-2/15* for a sphere (*e=0*) and goes towards 0 in the needle limit (*e=1*). Overall this means that all the $\Gamma_{ij}$ – factors are of the order 0.1 for our estimates below.

### Estimates and discussion
From Eq.(10) we can derive that for a good quality resonance in the polarizability the Raman scattering cross-section is proportional to the fourth power of the corresponding quality factor of the resonance (resonance frequency divided by damping). This is important since we found a very small value of $\Gamma_{ij} \cong 0.1$ above and we will find that $\delta_j$ is even smaller.

The general elastic properties of metals and the surface localization of the distorsion due to the vibration in itself implies that $\delta_j$ should be rather small. A simple estimate is $\delta_j \approx l_j/a_j^o$ where $l_j$ is the lattice constant in direction *j*, as a measure of the maximum deflection possible. Using the Lindeman melting criterion a 10% change would be too much for the system. We therefore take the conservative value $\delta_j \approx 0.01$ for our estimates in what follows. Notice also that if there is a relative stretch $\delta_j$ which is the same in all directions as for a breathing mode, however large amplitude it has, our theory conforms with the earlier findings that this mode is not Raman active expressed as $\sum_i \Gamma_{ij} = 0$.

We now have the ingredients necessary for estimating the cross-section. With a Drude dielectric function characterized by plasma frequency $\omega_p$ and scattering rate $\delta$ we can write

$$\sigma_R = (\omega_{inc}/c)^4 \left| \alpha_{ii}^2 \sum_j \Gamma_{ij} \delta_j / V \right|^2 \approx 10^{-6} V^2 (\omega_p^2/c\delta)^4 \approx 10^{-6} V^2 / l_o^4 \qquad (15)$$

where in the last line we have introduced a reasonably good resonance of quality factor 10 (to account for surface scattering bringing in the size of the system, otherwise its bulk value would be around 100). The electromagnetic length-scale $l_o \equiv c\delta/\omega_p^2 \approx 100 nm$. A particle of typical dimension *10nm* would then have a cross-section of the order of $10^{-22} m^2$. This is a value which carries a positive possibility to explore given typical Raman cross-sections for molecules. The problem is of course that the vibrations are in the meV range which makes it difficult for a Raman spectrometer to see the difference between the inelastic Raman and the elastic Rayleigh scattering. On top of this we have the traditional problems with small metal particles with respect to size distribution and interactions with substrate or between the particles. Using our model estimate we have the following ratio for Raman to Rayleigh

$$\sigma_R / \sigma_{Ray} = \left| \alpha_{ii} \sum_j \Gamma_{ij} \delta_j / V \right|^2 \approx 10^{-6} (\frac{\omega_o}{L\delta})^2 \approx 10^{-3} \qquad 16)$$

With the quality factor of the order of *10*. Thus the Raman signal is within the elastic signal and it is also three orders of magnitude smaller. However the experimental techniques in the area, see for instance reference 21, give hope that such a signal can be detected and shed light on the possible enhancement of a Raman signal due to the localized surface plasmons excited on the particle.

# Acknowledgement
S P Apell acknowledges hospitality and support from CFM/DIPC, San Sebastian, Spain.

# Appendix

In order to evaluate the derivate of the polarizability with respect to the direction of a deformation with a volume change of the particle we write

$$\sum_j q_j \frac{\partial \alpha_{ii}}{\partial q_j}\bigg|_0 = \frac{\partial \alpha_{ii}}{\partial V}(\sum_j \frac{\partial V}{\partial a_j} q_j \frac{\partial a_j}{\partial q_j}\bigg|_o) . \tag{A1}$$

The factor within parenthesis can be dealt with first. Using Eq.(8), the last derivate in Eq.(A1) can be combined with $q_j$ to form $a_j \delta_j$ for the result:

$$\sum_j q_j \frac{\partial \alpha_{ii}}{\partial q_j}\bigg|_0 = V \frac{\partial \alpha_{ii}}{\partial V} \sum_j \delta_j . \tag{A2}$$

To get to this we also used that $a_j \partial V / \partial a_j = V$ for the spheroids we are considering. Thus in what follows it is enough to pay attention to the volume derivate of the polarizability. In principle we have three contributions:

The *first* is the explicit volume dependence in the polarizability itself whereby $\partial \alpha_{ii} / \partial V = \alpha_{ii} / V$. It is seen to be of first order in the polarizability as compared to the treatment in the main body of the paper. Close to resonance we thus expect this contribution to be small in comparison, all else being equal.

The *second* contribution comes from a volume dependence in L itself when we include radiation damping since this is known to be important when particles get larger. For a simple estimate we do the customary replacement $\alpha \rightarrow \alpha/(1-2ik^3\alpha/3)$ [8] where k is the wave number of the incoming radiation. We then find that the result above is reduced even further by a factor $(1-2ik^3\alpha/3)^2$.

The *third* contribution is the possible dependence a volume change has on the plasma frequency of the free electron gas. To this end we get (L being the geometric factor in direction i):

$$\frac{\partial \alpha_{ii}}{\partial V} \equiv L(\alpha_{ii}/V)^2 \tag{A3}$$

It is of second order in the polarizability as the main contribution in this paper and in size it is L as compared to $\Gamma$, i.e. of the same order of magnitude.